# Network Activities Recognition and Analysis Based on Supervised Machine Learning Classification Methods Using J48 and Naïve Bayes Algorithm

Huang Fan, WKWSCI, NTU, Singapore, fhuang004@e.ntu.edu.sg

*Abstract*—Network activities recognition has always been a significant component of intrusion detection. However, with the increasing network traffic flow and complexity of network behavior, it is becoming more and more difficult to identify the specific behavior quickly and accurately by user network monitoring software. It also requires the system security staff to pay close attention to the latest intrusion monitoring technology and methods. All of these greatly increase the difficulty and complexity of intrusion detection tasks. The application of machine learning methods based on supervised classification technology would help to liberate the network security staff from the heavy and boring tasks. A finetuned model would accurately recognize user behavior, which could provide persistent monitoring with a relative high accuracy and good adaptability. Finally, the results of network activities recognition by J48 and Naïve Bayes algorithms are introduced and evaluated.

## I. Focused Forensic Problem

Given three datasets that do not have user activity label, the possible activities are browser using, music playing, and troubleshooting. And the user activity labels of three datasets are mutual exclusive, which provides more ways to make classification prediction task after the confirmation of one dataset label. However, this paper would focused on providing a general and reliable method to make classification on network activities.

## II. Survey of Related Work

### A. Academic Researches

Researchers have done many works on the application of machine learning method to network intrusion detections. For example, the usage of genetic algorithms and decision trees to create rules for intrusion detection expert system is a very nice implement of machine learning method. The rule generation is an impressive direction[1]. However, rule generation possess a bigger time complexity, which would possible lead to long response time and even memory crash that are fatal to some certain intrusion detection situations.

The supervised and unsupervised machine learning methods could be both useful during classifying user activities form network traffic. The combination of K-Means and Random Forest algorithms could obtain 97.37% accuracy[2].

Theoretically, the unsupervised algorithms (like K-Means and GM) and the supervised algorithms (like SVM and random forest) could be combined to produce a better accuracy than the models only use supervised algorithms. Because the pure supervised models would process the risks of losing flow information and low quality of training dataset labeling process.

### B. Open Source Projects

The main purpose of this paper is to explore and provide a solid, compact, easy-to-deploy, easy-to-adjust system for intrusion detection researchers and cyber security professionals. Therefore, the open source community is also an important reference and guidance for the design, development, and evaluation of this paper.

One machine learning method focused on the detection of attacks to IoT networks uses random forest and feed-forward neural network to classify the sample dataset. It labels the data as normal and malicious and then classify the test data with different types. The accuracy is 88%[3], while the project still possess the common shortage like huge memory consumption. This project show a good example of how to conduct machine learning method using Python. And it also demonstrates that the most common and important problem when conducting a machine learning system is how to relatively obtain the best accuracy or performance under the restrictions of resources consumption.

## III. Selection of Existing Computational Methods

The purpose of this project is mainly to build a compact machine learning system used to recognize the user activities with the best flexibility of machine learning algorithms and running environment. So, it is good to choose the Weka and the Wireshark as the main tool stack for this project.

As for the specific algorithm, I would choose to set decision tree type algorithm outcome as the baseline, and then make prediction based on the outcome. After that, I would choose to use different ways to finetune prediction outcome.

### A. Weka

The full name of the Weka is Waikato environment for knowledge analysis. It is a free, non-commercial, and open-sourced software for machine learning or data mining tasks, which is based on Java environment. The application with GUI and its Jar package could be downloaded from its official website. It supports various platforms including Windows, Mac, and Linux. The official instruction video and document is accessible in the official website. The main developer of Weka is from the University of Waikato of New Zealand.

The Weka has a large scale of algorithms clustered by hierarchical structures including data preprocessing, classification, clustering, association rules and visualization functions, which is a very helpful tools for low memory consuming tasks or small range testing tasks. The built-in visualization toolkits and detail parameter setting interfaces make it much easier and straightforward for users to operation. The user friendly interface and auto reminders would also help beginners to master this software even more faster.

In addition, users could import the official Jar package to the java programs and then they could use the exact same function and operations with all parameter settings just like those in Weka software. The Weka also support mainstream data science tools such as R, Python, and Spark for more integrated developments and researches.

*B. WireShark*

The Wireshark is not an intrusion detection system (IDS), because it will not alert or prompt any abnormal traffic behavior on the network. Consequently, the users should make careful analysis of the packets retrieved. The Wireshark does not modify the content of network packets, it only reflects the packet information in circulation. And it does not send packets to the network, in other words, it is a pure observe and record tool for internet activities. And this could let users to retrieve the raw and complete data directly from internet driver of the current computer. Then, they would have a better understanding the details and complete processes of network behavior.

As for the attributes of its retrieved data, the ID number and the time attribute shares basically the same function in the whole dataset. The time scale depends on the system time of the bag grabbing computer. So, the time attribute possesses more useful information than the ID number does. The deleting of ID number could be a good preprocessing method to improve the time consumption problem when training, validating, and testing.

Meanwhile, it is important to note that the time scales of the first and last messages captured are not highly accurate, which means that in the final evaluation users should take the negative influence from time attribute to final prediction result seriously.

Besides, users could also use the powerful filter engine of the Wireshark to filter out useful packets and eliminate the interference of irrelevant information. This is a very good way to generate the specific application network activities, which could contribute to improve the accuracy of trained model when making predictions.

*C. Decision Tree baseline*

The decision tree is a supervised learning, which uses decision analysis method to get and compare the probability of different leaves and nodes. And then evaluate and the best model with highest feasibility. It is a graphic method using probability analysis to provide an intuitive model format, which means that it is easy to generate and understand. It uses the entropy to measure and evaluate the degree of system clusters using algorithms like ID3, C4.5 (which is also known as J48), and C5.0. It is a very common classification method, and is suitable to be the baseline of classification tasks.

ID3 algorithm is based on information theory, and takes information entropy and information gain degree as the measurement standards, so as to generate the inductive classification of data.

J48 algorithm is a classification decision tree algorithm, which is based on the ID3 algorithm. This algorithm inherits the advantages of ID3 algorithm, and improves ID3 algorithm in the following aspects:

- Use the information gain rate to select attributes, which overcomes the disadvantage of the preference of selecting attributes with more values only using the information gain;
- Carry out pruning in the process of tree construction;
- Possess the capability of completing the discretization operation to continuous attributes;
- Possess the capability of handling incomplete data.

The J48 algorithm has the advantages of the highly readable classification rules and high accuracy. However, it also has the disadvantages of relatively higher time consumption when scanning and sorting the dataset and higher system I/O resources consumption compared with other decision tree models.

The high accuracy is more important than the resources and time consumption, so the J48 decision tree model would be conducted as the baseline of this project.

*D. Formulas*

The ID3 algorithm uses the information gain (IG) as the measure parameter and criteria during optimizing decision attributes, while the J48 algorithm uses the SplitInfo and the information gain rate (IGR) as the measure parameter and criteria. The IG, SplitInfo, and IGR could be calculated by the following formulas:

$$IG(attr) = Entropy(S) - Entropy(S|attr) \quad (1)$$

$$SplitInfo_{attr}(S) = -\sum_{v=v_0}^{v_n}\left(\frac{|S_v|}{S} + log_2 \frac{S_v}{S}\right) \quad (2)$$

$$IGR(attr) = \frac{IG(attr)}{SplitInfo(attr)} \quad (3)$$

Specifically, the J48 algorithm uses SplitInfo to standardize the information gain. If one attribute has more possible values, then its SplitInfo would be bigger. Finally, the J48 algorithm would make final decisions be lesser influenced by attributes simply owns more possible values and get a more accurate decision tree.

## IV. DATA GATHERED

In the machine learning process of this project, three kinds of datasets should be prepared.

### A. Training dataset

There are basically two ways to obtain the training dataset. Firstly, researchers and professionals could get the PCAP format files using the Wireshark software to fetch the network activities at the localhost. Secondly, they could also download the existing network activities records from some online platform or some open-sourced projects. However, it is a practical and convenient way to search for the PCAP format files from that are also generated by the Wireshark software when choosing the second way to get the training dataset.

To make the system and the approaches in this paper more flexible and adaptive, I choose to obtain the training dataset using the Wireshark to record the network activities at the localhost.

- Download and install the Wireshark software from its official website and follow the official instructions to complete the installation;
- No need to download other supplementary software from the other websites even if you have found some recommendation in others previews blogs;
- Open the Wireshark software, choose the network driver that the localhost computer or server are using;
- Start monitoring and recording the network activities.

There is one more feature of Wireshark need to know is that its outputs do not have labels but just raw records passing through the designated network hardwire, which could conflict with the supervise machine learning requirements. So it is necessary to confirm the recorded activities are easy to add classification labels, which means that users should only perform one specific kind of network activity to obtain useful training dataset during monitoring the network activities.

In this paper, the possible prediction results are browser using, music playing, and trouble shooting.

*1) For browser using:* Use only browser while the network is been monitored.

*2) For music playing:* Use only music player while the network is been monitored.

*3) For trouble shooting:* Use only terminal to run trouble shooting commands while the network is been monitored.

### B. Validation dataset

The Weka provide different kinds of separation or combination of training data and validation data.

- Use training set: validation dataset is exactly the same as the training dataset by default;
- Percentage split: separate the input training dataset in to two part, and the percentage of validation dataset is 33% by default and could be changed into any given percentage;
- Cross-validation: use the k-folds method to select the validation dataset, this is the chosen validation dataset setting in this paper;
- Supplied test set: could choose any local file or the URL file as the dataset, this is the chosen prediction method in this paper.

### C. Test dataset

The test datasets are provided in three separate files without classify label, which format is identical to the Wireshark records.

## V. DESIGN OF EXPERIMENTS

The experiment design is listed below:

### A. Configuration of the environment

Install and configure the Wireshark and Weka software and dependencies.

### B. Generate the training dataset

Generate the three kinds of datasets at localhost and export the records to the CSV format using the built-in export and convert function in the Wireshark.

Using CSV format is because this format separates the data with comma, which means the file is very simple and convenient for the further process. Besides, the CSV file could be edit directly through Microsoft Excel while still remains its comma separator. That would make the modify and labeling processes become very easy and fast.

### C. Import the datasets into the Weka to start training

Changing the CSV format datasets to the ARFF format datasets to make them readable to the Weka using the built-in tools of Weka.

### D. Data preprocess

Adjust and preprocessing the datasets to get better classification outcome.

### E. Apply classifier for training

Use J48 as the baseline classifier.

Use other classifiers to find the model with highest accuracy and the best prediction result.

### F. Preprocessing the splited test dataset

Generate three sub-datasets with three different labels for each test dataset and then test them using the generated model to find the right classification.

### G. Evaluation

Find out the best machine learning method based on the environment and test data. Then find the most reliable predictions of three test datasets.

## VI. ALGORITHM PREPARATION

This part is mainly about data preprocessing choices and the training outcome of three different classification algorithms.

## A. Data preprocessing

*1) Reduction of dataset dimension :* Delete the ID number column because the information it carries is basicly the same as the time column. Besides, reduction of the dimension would also greatly help to reduce the time and resource consumption when training and predicting.

*2) Delete the double quotation in the info column:* The double quotation marks would cause train fail because the programming language would identify the info column into serveral column and throws exception finally.

*3) Add labels for the train datasets and aggragate three datasets into one :* The supervised machine learning method needs labeled training dataset.

## B. Training outcome of different classification algorithms

The training process is conducted through three kinds of algorithm. The basic accuracies of them are listed below:

TABLE I. ALGORITHM TRAINING ACCURACY 10-FLODS

| Outcomes | Algorithm training using 10-folds validation | | |
|---|---|---|---|
| | Naïve Bayes | Naïve Bayes Updateable | J48 |
| Accuracy | 0.883028 | 0.883028 | 0.903668 |
| Run time | 0.1 s | 0.02 s | 0.77 s |

These three algorithm all achieved the accuracy above 0.85, which means that they are all reliable classifiers.

## VII. EVALUATION

### A. Algorithms

The detailed outcomes of three different algorithms are listed below:

TABLE II. ALGORITHM TRAINING 42K INSTANCES

| Outcomes | Algorithm training using 42198 instances | | |
|---|---|---|---|
| | Naïve Bayes | Naïve Bayes Updateable | J48 |
| Accuracy | 0.883028 | 0.883028 | 0.903668 |
| Run time | 0.1 s | 0.02 s | 0.77 s |
| Precision | 0.927 | 0.927 | 0.901 |
| Recall | 0.883 | 0.883 | 0.904 |
| F-measure | 0.894 | 0.894 | 0.902 |
| ROC Area | 0.984 | 0.984 | 0.987 |

After comparations of those three algorithms above, it is obvious to conclude that all of them perform pretty good in the training dataset. Their accuracy are around 90% after the 10-folds validation, and their ROC area are higher than 98%. There could be two reasons, one is that the algorithms performs very well in and this dataset is reliable and comprehensive to all kinds of situations could possibly exist. Another one is that the algorithm is cheated by the simple dataset.

The detailed comparison of three different algorithms are listed below:

*1) For Naïve Bayes algorithm:* This algorithm possesses a relatively low accuracy but a good time consumption, which means it is a good way to make classification tasks about network activities recognition.

*2) For Naïve Bayes Updateable algorithm:* This algortrm possesses the same good outcomes just like Naïve Bayes algorithm and have a even lower time consumption, which means it is a better choice to use this algorithm than the original Naïve Bayes algorithm.

*3) For J48 algorithm:* This algorithm possesses the best accuracy and ROC area but the worst time consumption (around 33 times of the Naïve Bayes Updateable algorithm) among three good algorithms.

In conclusion, the Naïve Bayes Updateable algorithm and The J48 algorithm are both excellent choices for network activities recognition tasks. However, the former one takes much lesser time consumption, while the latter possesses a better accuracy.

Consequently, if the users want a more accurate model, the J48 is an excellent choice, but they have to bear the huge time consumption. If they want a faster model with a quite good performance, the Naïve Bayes Updateable algorithm could be a better choice.

### B. Predicitons of classification

For higher accuracy, the J48 algorithm would be chosen to perform the prediction task.

TABLE III. PREDICITON POSSIBLITIES

| Outcomes | The J48 algorithm test outcome | | |
|---|---|---|---|
| | Sample 1 | Sample 2 | Sample 3 |
| Browser using | 0.459 | 0.139 | 0.521 |
| Music playing | 0.548 | 0.243 | 0.501 |
| Trouble shooting | 0.322 | 0.887 | 0.443 |

According to the prediction result, it is clear that the sample 2 would very likely to be the trouble shooting dataset. However, the prediction results of sample 1 and sample 3 are not very reliable considering the relatively low accuracy of the prediction from trained model. So, it is objective to conclude that the sample 1 has a higher possibility to be the browser using dataset and the sample 3 owns a relatively higher possibility to be the trouble shooting dataset.

The possible error analysis and the methods to improve the prediction results are elaborated at the following part.

### C. Possible error analysis

The system errors is generated by the built-in features of the research tools, incomplete test dataset or even the software or hardware features. For example, the different programming languages, the specific statistical features in the test dataset, or the differences between compiling and running environments (more specifically, the same machine learning program with fixed random seed would receive different accuracy when running in the CPU mode and GPU mode).

The experimental error is generated by the imprecise and incomplete experiment procedure design or the small coding errors during programming.

*1) The music player differences :* In generation part of the music playing dataset, I used the Spotfiy and the Netease Music to generate the training dataset.

*a) Operation method differences:* Unlike browsers owns the mainly similar network operations, the different music players could likely to have largely different network activities.

*b) Application design similarity:* The music player nowadays have more and more complicated and comprehensive functions and features, user could use the application to listen music, interacte with friends, post and view photos, watch videos and so on. Some music players even have their own built-in browsers, which makes it even more difficult to make classify prediction between browser using and music playing.

*2) The browser dataset generation :* The browser activities is various and hard to distinguish from other active softwares or system services. Use firewall to disconnect all other internet activities is too complicated and hard to fully accomplished.

*D. Possible ways to get better performance*

*1) The music player differences :*

*a) Operation method differences:* Monitor the network activities of the Spotify and the Netease Music separately and make comparison between the detailed differences between two network activity patterns. Could use mutiple clustering machine learning methods to identify if there is huge variances and to what degree the varances are. But the clustering method is a blackbox experiment with not much interpretability, there still ramains a lot work and exploration to be done.

*b) Application design similarity:* Could use a pure music player as a solution. But there is no such a thing among mainstream players. If it is a samll open-source software, then the comparison process still need to be done.

*2) The browser dataset generation :*

*a) Get the PID of the browser:* Use the Activity Monitor of Mac to get the the PID allocated to the browser (for example, the chrome)

*b) Get the port of the browser network interactions:* Use the terminal and command line '*netstat -anvp tcp |grep [PID]*'.

*c) Or set the specific site (like google) for browser reaching:* Use ping in the terminal to get the ip address of 'www.google.com' (like 172.217.194.105)

*d) Use filter function in the Wireshark to collect sepcific records:* For example, use 'ip.dst==172.217.194.105' to record only the network activities between localhost to specific ip. At the same time, use chrome to browse only on the google host website

*e) Shortcomings of this solution:* the collected dataset could be hightly homogeneous, which would very likely to weaken the accuracy of trained model when facing different test datasets.

*E. Possible improvement directions and ideas*

*1)* Fetch more network activity records to generate more reliable and comprehensive train dataset.

*2)* Try to use specific datasets of browsers and music applications collected by the professionals or on other environments.

*3)* Try more classification methods and then make cross comparison between different algorithms prediction results.

*4)* Conduct more data preprossessing methods to obtain a more accurate prediction result.

## Conclusion

Among three mentioned classification algorithms, if the users want a more accurate model, the J48 is an excellent choice, but they have to bear the huge time consumption. If they want a faster model with a quite good performance, the Naïve Bayes Updateable algorithm could be a better choice.

According to the prediction result, it is clear that the sample 2 would very likely to be the trouble shooting dataset. However, the prediction results of sample 1 and sample 3 are not very reliable considering the relatively low accuracy of the prediction from trained model. So, it is objective to conclude that the sample 1 has a higher possibility to be the browser using dataset and the sample 3 owns a relatively higher possibility to be the trouble shooting dataset.

## References

[1] C. Sinclair, L. Pierce and S. Matzner, "An application of machine learning to network intrusion detection," Proceedings 15th Annual Computer Security Applications Conference (ACSAC'99), Phoenix, AZ, USA, 1999, pp. 371-377, doi: 10.1109/CSAC.1999.816048.

[2] Labayen, V. (2020). Online classification of user activities using machine learning on network traffic. Computer Networks (Amsterdam, Netherlands : 1999), 181. https://doi.org/10.1016/j.comnet.2020.107557

[3] Harshilpatel1799. Iot-Cyber-Security-with-Machine-Learning-Research-Project.https://github.com/harshilpatel1799/Iot-Cyber-Security-with-Machine-Learning-Research-Project .2020

[4] Japkowicz, N., & Shah, M. (2011). Evaluating learning algorithms : a classification perspective / Nathalie Japkowicz, Mohak Shah. Cambridge University Press.